# Homogenization procedure for a metamaterial and local violation of the second principle of thermodynamics

**Nadia Mattiucci, Giuseppe D'Aguanno, Neset Akozbek, Michael Scalora, and Mark J. Bloemer**
*Charles M. Bowden Research Facility, RDECOM Bldg. 7804, Redstone Arsenal, AL 35898*

**Abstract**

Classical theory of crystals states that a medium to be considered homogeneous must satisfy the following requirements: a) the dimension of the elementary cell must be much smaller than the incident wavelength; b) the sample must contain a large number of elementary cells, i.e. it must be macroscopic with respect to wavelength. Under these conditions, macroscopic quantities can be introduced in order to describe the optical response of the medium. We analytically demonstrate that for a symmetric elementary cell those requirements can be relaxed, and it is possible to assign a permittivity and a permeability to a composite structure, even if the metamaterial cannot be considered homogeneous under the requirements stated above. However, the effective permittivity and permeability in some cases may give rise to unphysical, effective behaviors inside the medium, notwithstanding the fact that they satisfy requirements like being Kramers-Kronig pairs, for example, and are consistent with all the linear properties outside the structure (i.e. reflection, transmission, and absorption at all frequencies). In some situations the medium is assigned a magnetic response even though the medium is not magnetically active. In particular, we demonstrate that the homogenization procedure can lead to a medium that locally violates the second principle of thermodynamics. We also show that, in the non-homogeneous regime, it is not possible to describe the nonlinear behavior of the structure using an effective parameters approach, despite the possibility to assign an effective linear refractive index.

**I-Introduction**

The possibility to assign effective properties to composite structures has always captured the interest of scientists [1-19]. After the explosion of photonic band gap (PBG) research, starting almost 20 years ago, many researchers tried to assign an effective index to those artificial crystals [1-6]. More recently, after the astonishing growth of the field of negative index materials (NIMs) and meta-materials (MMs), the subject seems to have taken new life [7-19]. MMs are artificially constructed macroscopic composites made of periodically arranged inclusions. These inclusions, within a host medium or on its surface, are the building blocks of MMs and can be designed in order to produce material properties not obtainable with ordinary materials. The definition of MMs thus generally includes different classes of artificial materials including PBG structures, which have been extensively studied in the past, and NIMs. In particular NIMs have been shown to possess remarkable new properties not attainable in common materials like negative refraction [21], backward phase velocity, and super-lensing [22]. NIMs require both the dielectric permittivity and the magnetic permeability to be simultaneous negative. There are no known materials with such properties; in particular it is possible to obtain a negative permeability only with specifically designed inclusions. NIMs have been demonstrated in the microwave using split ring resonator (SRR) and conducting wires [21] (the SRRs account for the negative magnetic response while the wires are responsible for the negative electric response) and extended towards the optical regime using nanorods [23] or fishnet structures [24]. In order to characterize such composite structure it is useful to assign effective parameters. Moreover, the description of a composite structure with effective parameters may provide a simple physical understanding of the overall response of the structure. While, ideally those inclusions should be much smaller then the wavelength in current fabricated structures the inclusions are only a fraction of the incident wavelength, so there is still much debate on the possibility, limits and the opportunity for an effective description of such complex structures. In particular, the issue raised in regard to the possible violations of the second principle of thermodynamics [25-29] is very intriguing. Of course, several effective indexes can be defined for the same structure, depending on the approach followed in order to retrieve those effective properties. Some effective indexes are defined in order to explain the behavior inside the structure [2], others are more focused on the outside properties [1, 7, 11, 14]. In reference [3] effective properties are extracted in order to study scattering problems. Most papers study the effective index in connection to a single interface problem, i.e. the boundary between a common material (usually air) and a semi-infinite MM [9, 10, 15, 17]. In this case the effective parameter retrieval technique involves the definition of an effective impedance, that must match the actual surface impedance at the boundary between the two semi-spaces. Some retrieval techniques are based on the analysis of the Bloch diagrams for infinite periodic structure [2, 3, 9, 15, 17], others on field averaging [13, 19], and on transfer matrix [1, 5, 7, 11, 14]. In Ref. [12] the authors retrieve the effective indexes using two techniques, one based on the Bloch diagram (phase velocity), and the other on the transfer matrix (inversion method). We will focus our analysis on 1-D finite composite structures, and all the calculation will be performed according to the transfer matrix technique [30, 31]. Despite their simplicity, 1-D structures exhibit interesting effects like super-resolution [32-33] and negative refraction [34-36]. Moreover, the analytical theory developed for 1-D structures is often used to explain, at least qualitatively, the behavior of more complicated 2-D and 3-D structures. We would like to point out that the effective parameters analyzed in this paper are defined starting from the connection between the fields at the input and output interfaces: i.e. we use an intrinsic black box approach. In other words, no physical assumptions are made on the particular nature of structure and the effective parameters are extracted using the scattering data (reflection and transmission) of the structure. This retrieval technique, that we can refer to as "parameter extraction" has been widely used to retrieve the effective properties (permeability and permittivity) of fabricated MMs [7, 11, 14]. These "effective parameters" should not be confused with parameters resulting from effective medium theories (for example, theories developed to describe a composite consisting of metal particles in a glass host such as the Maxwell Garnett theory [37]). The paper is organized as follows: in Section II we discuss a general theory based on the transfer matrix technique in order to retrieve the effective parameters for a 1-D multilayer structure, in Section III we give some examples of the application of the technique to particular structures, in Section IV we discuss the local violation of the second principle of thermodynamics, and finally in Section V we discuss the possibility to assign effective nonlinear parameters to the structure.

**II-Periodic Multilayers: General theory**

Let us start by considering an elementary cell of length $\Lambda$, which for the moment represents our entire structure. We assume that the elementary cell can be made of a generic combination of layered materials. A plane wave incident on that structure at a certain angle (on the y-z plane) can be written as $\vec{E} = E(z)e^{ik_y y}e^{-i\omega t}\hat{i}$ (electric field) for TE polarization, or as $\vec{H} = H(z)e^{ik_y y}e^{-i\omega t}\hat{i}$ (magnetic field) for TM polarization. The amplitude and phase of the field at the input and output surfaces are linked by a matrix relation:

$$\begin{pmatrix} \Psi_{Out} \\ \varphi_{Out} \end{pmatrix} = \begin{pmatrix} m_{11} & m_{12} \\ m_{21} & m_{22} \end{pmatrix} \begin{pmatrix} \Psi_{In} \\ \varphi_{In} \end{pmatrix} \quad . \quad (1)$$

For TE polarization, $\Psi=E$ and $\varphi=(1/\mu)dE/dz$. For TM polarization, $\Psi=H$ and $\varphi=(1/\varepsilon)dH/dz$. The transfer matrix of the elementary cell is simply given by the product of the transfer matrix of the layers that compound the elementary cell:

$$\begin{pmatrix} m_{11} & m_{12} \\ m_{21} & m_{22} \end{pmatrix} = \prod_{Elementary Cell} \begin{pmatrix} \cos(q_i d_i) & \frac{\alpha_i}{q_i}\sin(q_i d_i) \\ -\frac{q_i}{\alpha_i}\sin(q_i d_i) & \cos(q_i d_i) \end{pmatrix}, \qquad (2)$$

where $d_i$ is the thickness of the $i^{th}$ layer, $q_i$ is the longitudinal component of the k vector in the $i^{th}$ layer, $\alpha_i=\mu_i$ for TE polarization or $\alpha_i=\varepsilon_i$ for TM polarization, being $\mu_i$ and $\varepsilon_i$ the bulk magnetic permeability and the electric permittivity of the $i^{th}$ layer. The idea is to find the effective indices by simply setting the transfer matrix of our elementary cell equal to the transfer matrix of a single layer of the effective or equivalent medium, i.e.:

$$\begin{pmatrix} \cos(q_{eff}\Lambda) & \frac{\alpha_{eff}}{q_{eff}}\sin(q_{eff}\Lambda) \\ -\frac{q_{eff}}{\alpha_{eff}}\sin(q_{eff}\Lambda) & \cos(q_{eff}\Lambda) \end{pmatrix} = \begin{pmatrix} m_{11} & m_{12} \\ m_{21} & m_{22} \end{pmatrix}. \qquad (3)$$

In Eq. (3), $\alpha_{eff}=\mu_{eff}$ (effective permeability) for TE polarization, or $\alpha_{eff}=\varepsilon_{eff}$ (effective permittivity) for TM polarization. $q_{eff}$ is the longitudinal component of the k vector in the effective medium: $q_{eff} = k_{z,eff} = n_{eff}\sqrt{(\omega/c)^2 - (k_y/n_{eff})^2}$, where $n_{eff} = \pm\sqrt{\varepsilon_{eff}\mu_{eff}}$ is the effective index. The sign in front of the square root must be chosen in a way that ensures the Poynting vector refracted inside the effective medium will be directed away from the interface into the refracting material itself. Eq. (3) provides 4 equations for 2 unknown quantities so, as already argued in ref. [8], in general there is no solution. However, it is very simple to determine the general condition that allows a unique solution. Let us start by writing the transfer matrix of the structure in term of the transmission (t) and the reflection (r, $\tilde{r}$) coefficients [38]:

$$\begin{pmatrix} m_{11} & m_{12} \\ m_{21} & m_{22} \end{pmatrix} = \begin{pmatrix} \dfrac{(1-r)(1+\tilde{r})+t^2}{2t} & i\dfrac{\alpha_0}{q_0}\dfrac{(1+r)(1+\tilde{r})-t^2}{2t} \\ -i\dfrac{q_0}{\alpha_0}\dfrac{(1-r)(1-\tilde{r})-t^2}{2t} & \dfrac{(1+r)(1-\tilde{r})+t^2}{2t} \end{pmatrix}, \quad (4)$$

where, r is the reflection for left to right (LTR) incidence, $\tilde{r}$ is the reflection for right to left (RTL) incidence, and t is the transmission. For simplicity, we are assuming that the input and the output media are the same. The subscript 0, in Eq.(4), refers to the immersion medium. Eq. (3) becomes:

$$\begin{pmatrix} \cos(q_{eff}\Lambda) & \dfrac{\alpha_{eff}}{q_{eff}}\sin(q_{eff}\Lambda) \\ -\dfrac{q_{eff}}{\alpha_{eff}}\sin(q_{eff}\Lambda) & \cos(q_{eff}\Lambda) \end{pmatrix} = \begin{pmatrix} \dfrac{(1-r)(1+\tilde{r})+t^2}{2t} & i\dfrac{\alpha_0}{q_0}\dfrac{(1+r)(1+\tilde{r})-t^2}{2t} \\ -i\dfrac{q_0}{\alpha_0}\dfrac{(1-r)(1-\tilde{r})-t^2}{2t} & \dfrac{(1+r)(1-\tilde{r})+t^2}{2t} \end{pmatrix}. \quad (5)$$

It is clear that we can have a unique solution only if $\tilde{r}=r$, *i.e. the structure must be symmetric*, so that the diagonal elements are equal in both matrixes, and the equations arising from the off diagonal elements are equivalent (although the transmittance through a linear medium is always symmetric the reflection is not necessarily symmetric. As an illustration a partially reflecting mirror with an absorbing layer deposited on one side has a symmetric transmission but an asymmetric reflection). Therefore, in the case of a symmetric elementary cell, from equation 5, in accordance with Ref. [20], we derive:

$$\cos(q_{eff}\Lambda) = \dfrac{1-r^2+t^2}{2t}, \quad (6a)$$

$$\dfrac{\alpha_{eff}}{q_{eff}} = \pm \dfrac{\alpha_0}{q_0}\sqrt{\dfrac{(1+r)^2-t^2}{(1-r)^2-t^2}}. \quad (6b)$$

Note that at normal incidence and the immersion medium being vacuum $\dfrac{\alpha_{eff}}{q_{eff}} = \sqrt{\dfrac{\mu_{eff}}{\varepsilon_{eff}}}\dfrac{c}{\omega} = \pm\dfrac{c}{\omega}\sqrt{\dfrac{(1+r)^2-t^2}{(1-r)^2-t^2}}$ and eq. (6b) becomes: $Z_{eff} \equiv \sqrt{\dfrac{\mu_{eff}}{\varepsilon_{eff}}} = \pm\sqrt{\dfrac{(1+r)^2-t^2}{(1-r)^2-t^2}}$,

where $Z_{eff}$ is the effective impedance of the medium, in agreement with the results reported in reference 7 for a generic metamaterial.

Of course, any finite periodic structure based on the same symmetric elementary cell is characterized by the same effective properties. In fact, if we consider a generic N-period structure (of length $L=N\Lambda$), then for the transfer matrix of the periodic structure the following equivalence holds:

$$\begin{pmatrix} m_{11} & m_{12} \\ m_{21} & m_{22} \end{pmatrix}^N = \begin{pmatrix} \cos(q_{eff}\Lambda) & \frac{\alpha_{eff}}{q_{eff}}\sin(q_{eff}\Lambda) \\ -\frac{q_{eff}}{\alpha_{eff}}\sin(q_{eff}\Lambda) & \cos(q_{eff}\Lambda) \end{pmatrix}^N = \begin{pmatrix} \cos(q_{eff}N\Lambda) & \frac{\alpha_{eff}}{q_{eff}}\sin(q_{eff}N\Lambda) \\ -\frac{q_{eff}}{\alpha_{eff}}\sin(q_{eff}N\Lambda) & \cos(q_{eff}N\Lambda) \end{pmatrix}.$$

(7)

In other words, the N-period structure is equivalent to a bulk of the same effective medium of the elementary cell and having the same length $L=N\Lambda$ of the N-period structure itself: i.e. the dispersion properties of the single, symmetric elementary cell and the dispersion of the N-period structure made of the same elementary cell are identical.

As a direct consequence of equations (4) and (5), $q_{eff}$ (the longitudinal component of the k vector in the effective medium) results to be nothing more than the Bloch wavevector ($K_B$), in fact by definition $K_B \equiv \mathrm{ArcCos}((m_{11}+m_{22})/2)/\Lambda = q_{eff}$. It is worthwhile to expand upon the last statements. The simplest way to construct a 1-D PC is to alternate periodically two different dielectric layers: one of thickness $d_1$ and refractive index $n_1$, the second of thickness $d_2$ and refractive index $n_2$ as shown in figure 1a. The elementary cell of this infinite structure can be extracted in many different ways. In figure 1 we represent the most popular elementary cells: the two symmetric elementary cells (figure 1b and 1c) and the two asymmetric (figure 1d and 1e). In the infinite structure the difference between one elementary cell and another is just a matter of fixing the origin of the coordinate system, and choosing one or the other does not change the physical properties of the structure because obviously the medium has translational invariance. In the finite structure (especially for short structures, i.e made of few periods) the translational invariance is broken and different elementary cells result in different physical behaviors. Note that also the authors of Ref. [15] come to the conclusion that for finite structures the interfaces play a vital role and the physics of the system as well as the effective indexes depend on the truncation of the crystal. What we have shown here is that an effective wave-vector (Bloch vector) can be unambiguously defined for a finite structure containing an arbitrary number N of elementary cells if and only if the elementary cell is symmetric. In the case that the elementary cell is not symmetric the effective wave-vector defined by Eq.(6a) can be used to describe the properties of the structure only if N>>1. This finding suggests that some caution should be exercised when applying the effective parameter retrieval technique to metamaterials made of few, non symmetric elementary cells. As an example, in Fig.2 we show the wavevector dispersion calculated from Eq.(6a) for a symmetric elementary cell as described in Fig.1b and we compare it with the dispersion of two structures made respectively of N=1 and N=10 asymmetric elementary cells described in figure 1d. As one may expect, in accordance with the theory exposed above, the dispersion of the finite structures tends to the Bloch dispersion for N>>1. The fact that a periodic multilayer structure is equivalent, at least in the quasi-static limit, to a homogeneous and optically anisotropic single layer (at angular incidence the refractive index depends on the polarization) is a well-known result [31]. Unfortunately, older papers [1, 5, 6] fail to find the exact expression of the electromagnetic properties of the equivalent layer at high frequency (i.e. by high frequency we mean all the wavelengths shorter than a few tens of elementary cells). In Ref. [1] the authors did find a general expression, but in order to do so they considered the length of the equivalent layer as a free parameter, so that in the high frequency regime the length of the equivalent layer is different from the actual length of the structure. In order to overcome this limitation we have to accept the possibility that the effective medium is magnetically active even if the single constituents are not. The implication is that the effective medium model is giving an erroneous picture of the actual constituents that make up the medium, but if one is primarily concerned with only the input and output fields then this can be an acceptable compromise. It is worthwhile to note that in the low frequency regime all the homogenization procedures are consistent with each other, and they converge to the well known conclusion that a 1-D periodic structure is equivalent to a uniaxial crystal with the optical axis perpendicular to the stratification, as for example pointed out in Refs. [39, 40]. The ordinary permittivity and permeability are simply given by their spatial average (weighted arithmetic average) while the extraordinary electromagnetic parameters are given by the weighted harmonic average.

**III-Numerical examples**

Let us begin our analysis by considering for simplicity a 1-D Photonic Crystal with a symmetric elementary cell made of two different materials of index of refraction $n_1$ and $n_2$, with total thickness $d_1$ and $d_2$, respectively. The elementary cell will thus consist of a three layer structure with the first and the third layers being equal (see figure 1). This leaves two possible choices for the elementary cell: $n_1/n_2/n_1$ with thicknesses $0.5d_1/d_2/0.5d_1$ (figure 1b) and $n_2/n_1/n_2$ with thicknesses $0.5d_2/d_1/0.5d_2$ (figure 1c). Of course the same infinite structure corresponds to both elementary cells, but while the effective index will always be the same for both structures (the Bloch wave vector is the same), the effective permeability and the effective permittivity will differ at high frequencies. In other words, while the diagonal elements of the transfer matrix (calculated according to Eq. 2) are the same for both elementary cells, the off-diagonal elements are not equal. In figure 3 and 4 $n_{eff}$, $\varepsilon_{eff}$ and $\mu_{eff}$ are calculated for both symmetric elementary cells.

We will now analyze the behaviour of the periodic structures based on the elementary cell represented in figure 1a (details on the elementary cell are given in the caption of figure 1). In figure 3, the effective parameters ($n_{eff}$, $\varepsilon_{eff}$ and $\mu_{eff}$) are shown for normal incidence. The imaginary part of $\varepsilon_{eff}$ and $\mu_{eff}$ exhibit alternate resonant and anti-resonant behaviors, to confirm the fact that this kind of effect is linked to periodicity [25, 28]. Moreover, for all the indexes the real and imaginary parts are Kramers-Kronig (K-K) related. If we think of the structure as a black box, from the outside it is impossible to distinguish between the effective medium and the real structure. So that, those effective indexes can be used in order to predict or explain the linear behavior of the structure. Of course, they may not provide any physical insight on what is really happening inside the structure. In our case, for example, at high frequencies the electromagnetic fields distributions inside the effective medium can be radically different from the ones inside the real structure. This aspect can have important consequences, as we will see in the next Sections.

We would like to remark that, although in our examples we considered for simplicity only 1-D all-dielectric structures, our theory applies to any kind of stratification as well, including metallo-dielectric structures. 1-D metallo-dielectric structures with a symmetric elementary cell have been used to achieve broadband super-resolution [33], for example. Our theory suggests that for such kind of structures an effective permittivity and permeability can be unambiguously defined and all the properties outside the structure, including super-resolution, negative refraction and cloaking, can be perfectly retrieved by using the effective parameters.

**IV-Local violation of the second principle of thermodynamics**

There has been considerable debate about the dissipation of heat in the effective medium view, relating to possible violations of the second principle of thermodynamics [25-28]. Let us see how our simple dielectric stack behaves. The dissipated heat rate per unit length ($Q_z$) is given by [40]:

$$Q_z = \frac{\omega}{2}\left(\text{Im}(\varepsilon(z))|E_\omega(z)|^2 + \text{Im}(\mu(z))|H_\omega(z)|^2\right) \quad (8)$$

While, the total dissipated heat rate (Q) by a planar structure of length L is:

$$Q = \frac{\omega}{2}\left(\int_0^L \text{Im}(\varepsilon(z))|E_\omega(z)|^2 dz + \int_0^L \text{Im}(\mu(z))|H_\omega(z)|^2 dz\right) \quad (9),$$

The expressions of the local dissipated heat rate and total dissipated heat rate inside the effective structure, $Q_{z,\text{eff}}$ and $Q_{\text{eff}}$ respectively, are formally identical to Eqs. (8,9) except that effective quantities replace the actual ones:

$$Q_{z,\text{eff}} = \frac{\omega}{2}\left(\text{Im}(\varepsilon_{\text{eff}})|E_{\omega,\text{eff}}(z)|^2 + \text{Im}(\mu_{\text{eff}})|H_{\omega,\text{eff}}(z)|^2\right) \quad (10)$$

$$Q_{\text{eff}} = \frac{\omega}{2}\left(\text{Im}(\varepsilon_{\text{eff}})\int_0^L |E_{\omega,\text{eff}}(z)|^2 dz + \text{Im}(\mu_{\text{eff}})\int_0^L |H_{\omega,\text{eff}}(z)|^2 dz\right) \quad (11),$$

Where the effective electric and magnetic fields ($E_{\omega,\text{eff}}$ and $H_{\omega,\text{eff}}$) are the fields that one would expect to find in a homogeneous layer having $\varepsilon_{\text{eff}}$ and $\mu_{\text{eff}}$ as electric permittivity and magnetic permeability. The field distribution inside the real structure ($E_\omega(z)$ and $H_\omega(z)$) and inside the effective medium ($E_{\omega,\text{eff}}(z)$ and $H_{\omega,\text{eff}}(z)$) are calculated according to the transfer matrix theory for LTR incidence. In each homogeneous layer and in the left semi-space the fields are a superposition of two plane waves (incident and reflected field) while, on the right semi-space each field reduces to a plane wave propagating LTR (transmitted field). Those stationary solution of the Maxwell equations (known also as mode of the structure) can be defined for any finite structure and must not be confused with the Floquet-Bloch modes defined for periodic infinite structures.

In our example, $Q_{eff}$, is positive at each frequency, and is equal to the absorption of the structure; on the contrary, $Q_{z,\,eff}$ can show unphysical behaviors as also argued in reference [26]. In fact we find that at high frequency, there are regions inside the structure in which $Q_{z,\,eff}$ is negative (see figure 4), giving rise to local violation of the second principle of thermodynamics [41]. Once again we stress that the violation of the second principle of thermodynamics is local and not global, in fact the total dissipated heat rate $Q_{eff}$ is always positive while the local dissipated heat rate $Q_{z,\,eff}$ shows "cold" and "hot" spots. Note that $Q_{eff}$ exhibits negative values both inside the gaps and in the pass bands. The behavior of $Q_{eff}$ inside the gaps is linked to the periodicity, while the behavior in the pass bands is determined by the actual absorption of the materials composing the structure. If the structure is made of non absorbing materials the unphysical behavior of $Q_{eff}$ in the pass bands disappears. In figure 5 we compare the behavior of the real structure with the behavior of the effective medium at the frequency $\omega=0.56\omega_0$ (inside the gap). The real structure is behaving as expected: it is dissipating energy in the absorbing layers with a strength proportional to the field intensity. On the other hand, the effective medium, which is supposed to be homogeneous, is characterized by alternate hot and cold regions, as if it were dissipating and generating energy at the same time. It is not unusual that effective parameters can behave in a non physical way under some circumstances. For example in Ref [29] violation of the second principle of thermodynamics has been found for PCs with high index modulation.

**V-Effective nonlinear parameters**

Until now we have focused on the linear effective properties of our multilayer. We have assessed that it is possible to assign linear effective parameters to compound structures with a symmetric elementary cell and that even in the high frequency regime these linear parameters are able to recover the global properties of the structure, i.e. reflection, transmission, absorption, and moreover they are Kramers-Kronig pairs. We will now briefly investigate the possibility to assign nonlinear parameters too. Let us assume that one type of layer inside the stack has a nonlinear quadratic response ($d^{(2)}$). We seek an equivalent medium with an effective quadratic response ($d^{(2)}_{eff}$). In the undepleted pump approximation and for normal incidence of the pump field, the second harmonic (SH) field can be expressed in term of the pump field and the modes of the structure at SH [42]:

$$E_{2\omega}(z) = -\frac{(2\omega)^2}{c^2} \int_0^L G_{2\omega}(\zeta, z) d^{(2)}(\zeta)(E_\omega(\zeta))^2 d\zeta \qquad (12).$$

In eq. (12), $G_{2\omega}(\zeta, z)$ is the Green function that takes the following expression in term of the LTR ($\Phi^+_{2\omega}(z)$) and RTL ($\Phi^-_{2\omega}(z)$) modes of the structure at the SH:

$$G_{2\omega}(\xi, z) = \frac{1}{2ik_0^{2\omega} t_{2\omega}} \begin{cases} \Phi^-_{2\omega}(\zeta)\Phi^+_{2\omega}(z) & 0 \leq \zeta < z \\ \Phi^+_{2\omega}(\zeta)\Phi^-_{2\omega}(z) & z < \zeta \leq L \end{cases} \qquad (13).$$

Similar equations can be written for the effective fields in the effective medium. We impose, of course, that the SH field at the input surface (backward SH generation) and the SH field at the output surface (forward SH generation) must be the same for both the real and the equivalent structure, i.e. $E_{2\omega}(0) = E_{2\omega, eff}(0)$ and $E_{2\omega}(L) = E_{2\omega, eff}(L)$. These conditions allow us to derive two independent expressions for $d^{(2)}_{eff}$:

$$\frac{d^{(2)}_{eff,Forw}}{d^{(2)}} = \frac{\int_0^L grat(\zeta)\Phi^+_{2\omega}(\zeta)(E_\omega(\zeta))^2 d\zeta}{\int_0^L \Phi^+_{2\omega,eff}(\zeta)(E_{\omega,eff}(\zeta))^2 d\zeta}, \quad \frac{d^{(2)}_{eff,Back}}{d^{(2)}} = \frac{\int_0^L grat(\zeta)\Phi^-_{2\omega}(\zeta)(E_\omega(\zeta))^2 d\zeta}{\int_0^L \Phi^-_{2\omega,eff}(\zeta)(E_{\omega,eff}(\zeta))^2 d\zeta} \qquad (14)$$

In order to assign an effective electric nonlinearity to the structure, the two independent calculations should give the same result, which must be independent from the number of elementary cells considered. Unfortunately, this is true only in the low frequency limit where the effective nonlinearity is simply given by the spatial average of the nonlinearities (as common sense would suggest). In figure 6 we show, as a representative example, the real and the imaginary parts of $d^{(2)}_{eff,Forw}$ and $d^{(2)}_{eff,Back}$, calculated for the structure described in the caption of figure 1 for LTR incidence of the pump field. Outside of the low frequency regime, the two retrieved nonlinear parameters are remarkably different, and their imaginary parts are different from zero. Moreover, only the real and imaginary parts of $d^{(2)}_{eff,Forw}$ are Kramers-Kronig related.

In the high frequency regime, the results change with the number of elementary cells considered and the incident pump conditions. This result is not surprising, in fact from Eqs.(14) it is clear that the SH generation in the undepleted regime is strictly related to the behavior of the linear fields inside the structure, and we have already seen that at high frequencies the effective parameters do not recover that behavior.

Previous theoretical studies on second order nonlinear process in MMs [42-45] consider only one kind of nonlinearity at a time (an electric $\chi^{(2)}$ [43, 46] or a magnetic $\chi^{(2)}$ [44-46]). Our simple example suggests that it is unlikely that current MMs, in which the elementary cell is bigger than 1/10 of the incident wavelength, may be simply described by a homogeneous nonlinear parameter. It is also equally unlikely that experimental results like those presented in ref [47] can be described by such a theory. It may be possible to override at least some of these limitations by looking for an equivalent medium with simultaneous electric and magnetic nonlinearity, and/or by introducing different kinds of nonlinearities associated with the elementary cell, the periodicity and the number of cells leading to a more complicated theory.

**VI-Conclusions**

In conclusion we can state that (mathematically speaking) a linear structure is equivalent to an effective K-K medium (regardless of frequency) provided that the elementary cell of the structure is symmetric. The use of this retrieval technique may be questionable when applied to a MM whose elementary cell is strongly asymmetric. As shown in figure 1, in the case of an asymmetric elementary cell, the retrieved effective parameters depend on the number of periods especially for structures containing few periods. The equivalence is valid only outside the structure. Inside the structure and at high frequencies, the behavior of the actual fields can be remarkably different from the behavior of the effective fields. As a consequence, at higher frequencies, the equivalent medium shows local violation of the second principle of thermodynamic (the effective medium becomes unphysical) and the effective index picture can no longer be used in order to predict the linear behavior of the fields inside the structure. The equivalent medium is in general a magnetically active medium. Once again, we would like to caution the reader that this equivalence with an effective medium, which is magnetically active, must, of course, be intended only in a mathematical sense, and no information regarding the actual electric and magnetic properties of the real structure can be inferred from the equivalent structure. From a practical point of view, this means that a mere parameter extraction based on scattering data is not sufficient to characterize the MM and it must always be accompanied by a physical model that describes the internal behavior of the structure. In our example of a simple 1-D multilayer made of nonmagnetic materials it is clear that no magnetic character can be associated with the structure, on the other hand an imprudent interpretation of the results would have brought to the conclusion that the structure is magnetically active.

As regards the nonlinear properties, we come to the conclusion that a 1-D PC made of $\chi^{(2)}$ materials, in the low frequency regime, is equivalent to a uniform $\chi^{(2)}$ medium while, in the high frequency regime, it is not possible to keep such equivalence. Those results obtained analytically for 1-D structures can be used in order to interpret the super-resolution from 1-D metallo-dielectric structures [33], for example, and in general have important consequences also for more complex structures like current MMs, where an analytical theory is hard to develop and 1-D results are often extended [6-10] to describe their effective behavior.

**Acknowledgements**

N. M., G. D. and N.A. thank the National Research Council for financial support.

FIGURE CAPTIONS

Figure 1: a) Infinite 1-D PC made of two different layers. The first layer has thickness $d_1=0.250\lambda_0$ and refractive index $n_1=2+0.01i$. The second layer has thickness $d_2=0.125\lambda_0$ and refractive index $n_2=4$. Where, $\lambda_0$ is the reference wavelength in vacuum. b) Schematic representation of the symmetric elementary cell $n_1/n_2/n_1$, $0.5d_1/d_2/0.5d_1$: a three layer symmetric structure where, the first and the third layer are $0.5d_1$ thick and have refractive index $n_1$ while the second layer is $d_2$ thick and has refractive index $n_2$. c) Schematic representation of the symmetric elementary cell $n_2/n_1/n_2$, $0.5d_2/d_1/0.5d_2$: a three layer symmetric structure where, the first and the third layer are $0.5d_2$ thick and have refractive index $n_2$ while, the second layer is $d_1$ thick and has refractive index $n_1$. d) Schematic representation of the asymmetric elementary cell $n_1/n_2$, $d_1/d_2$: a two layer asymmetric structure where, the first layer is $d_1$ thick and has refractive index $n_1$ while the second layer is $d_2$ thick and has refractive index $n_2$. e) Schematic representation of the asymmetric elementary cell $n_2/n_1$, $d_2/d_1$: a two layer asymmetric structure where, the first layer is $d_2$ thick and has refractive index $n_2$ while the second layer is $d_1$ thick and has refractive index $n_1$.

Figure 2 (Color online): a) Real part of the effective wave-vector multiplied the length of the elementary cell at normal incidence. b) Imaginary part of the effective wave-vector multiplied the length of the elementary cell at normal incidence. The solid line (Bloch dispersion) refers to a structure made of one symmetric elementary cell as described in Fig.1b. The long-dashed line refers to a structure made of one asymmetric elementary cell as described in Fig. 1d. The short-dashed line refers to a structure made of 10 asymmetric elementary cells. d1 and d2 are respectively the total thicknesses of the materials $n_1$ and $n_2$ composing the elementary cells. The effective wave-vector is retrieved using the transmission and the reflection coefficients as in equation 6a of the main text. The shaded areas indicate the PC's gaps.

Figure 3 (Color online): a) Schematic representation of the symmetric elementary cell $n_1/n_2/n_1$ described in the caption of Fig.1b b) Real, imaginary and KK of the imaginary part of the effective refractive indexes. c) Real, imaginary and KK of the imaginary part of the effective electric permittivity. d) Real, imaginary and KK of the imaginary part of the effective magnetic permeability. The shaded areas indicate the PC's gaps.

Figure 4 (Color online): a) Schematic representation of the symmetric elementary cell $n_2/n_1/n_2$ described in the caption of Fig.1c b) Real, imaginary and KK of the imaginary part of the effective refractive indexes. c) Real, imaginary and KK of the imaginary part of the effective electric permittivity. d) Real, imaginary and KK of the imaginary part of the effective magnetic permeability. The shaded areas indicate the PC's gaps.

Figure 5 (Color online): a) Local (effective) heat dissipated $Q_{z,\,eff}$ as a function of frequency and position inside an effective medium of length $L=1.875\lambda_0$, equivalent to a 5-periods 1-D PC having as elementary cell the structure described in the caption of figure 1b. b) Dissipated heat per unit length inside a 5 period structure, $Q_z$, and inside the effective medium of the same length, $Q_{z,\,eff}$ for an incident wavelength of $0.56\omega_0$, where $\omega_0$ is the reference frequency.

Figure 6 (Color online): a) Real and b) imaginary part of the forward and backward effective second order nonlinearity. $\dfrac{d^{(2)}_{eff,Forw}}{d^{(2)}}$ and $\dfrac{d^{(2)}_{eff,Back}}{d^{(2)}}$ are calculated according to Eq. 14 of the main text for the five period structure described in the caption of Fig. 5. We are also assuming that the quadratic nonlinearity ($d^{(2)}$) is only in the layers of refractive index $n_1$.

Figure 1:

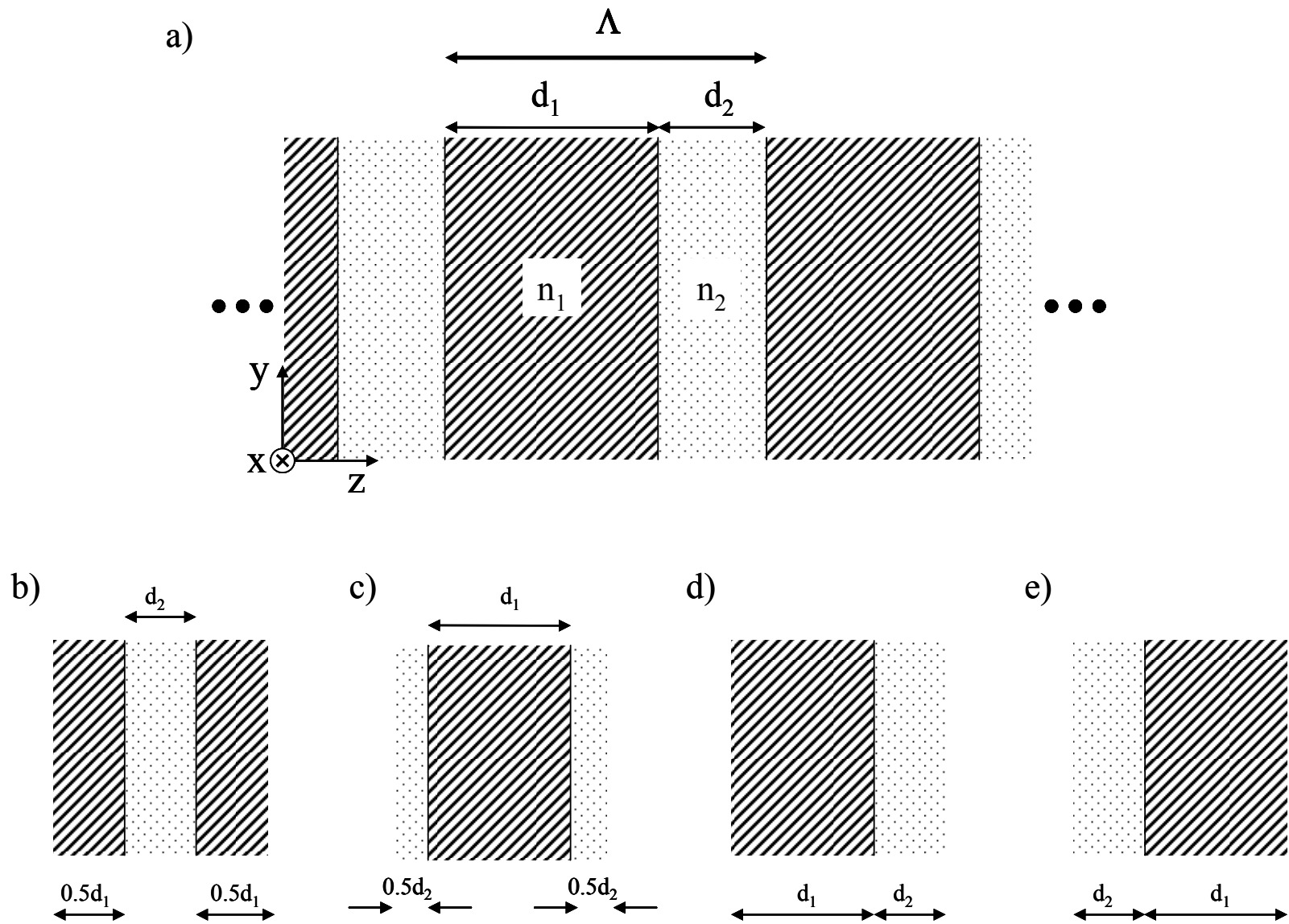

Figure 2:

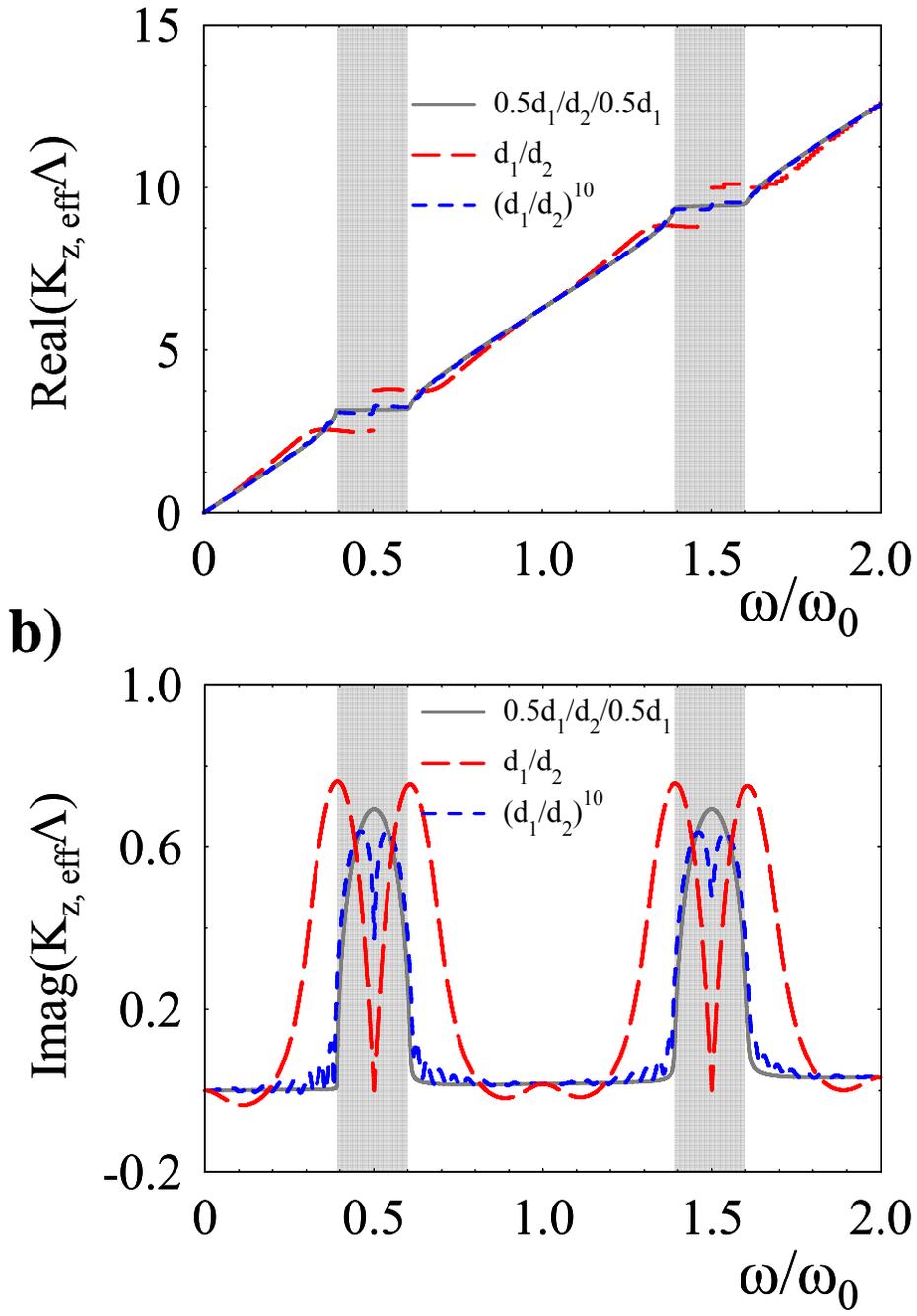

Figure 3:

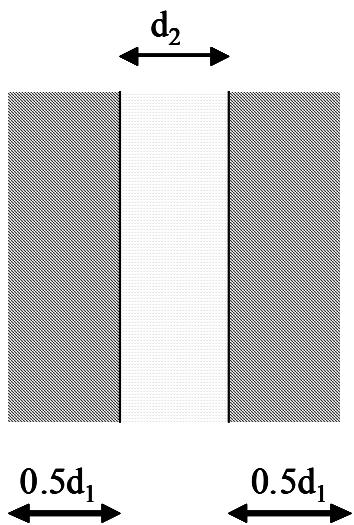
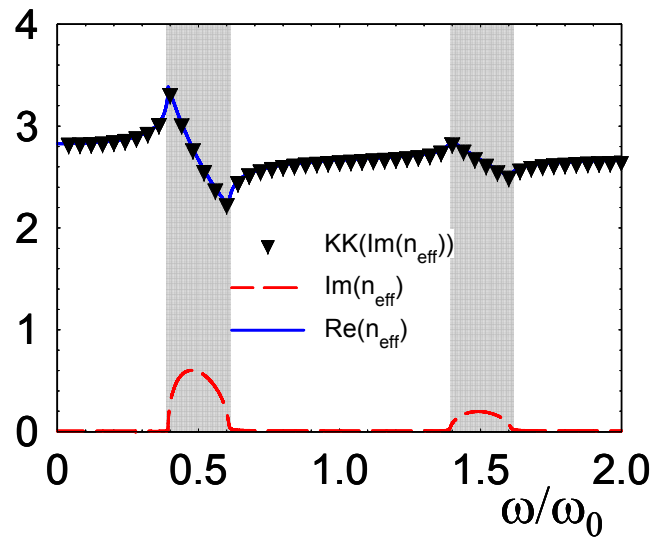
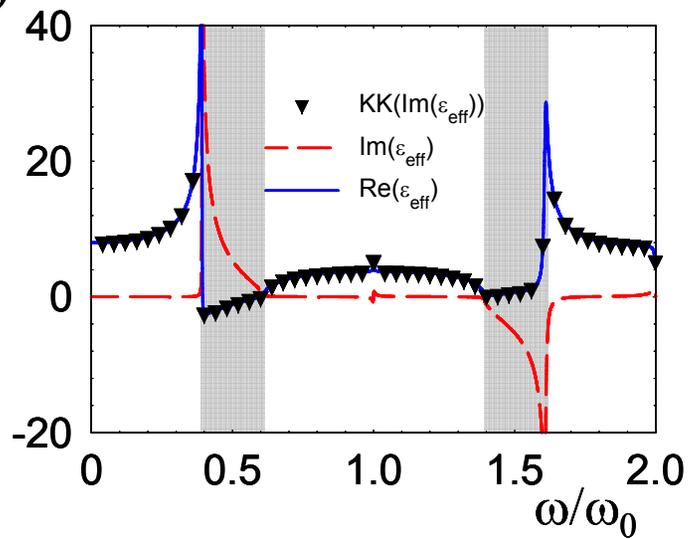
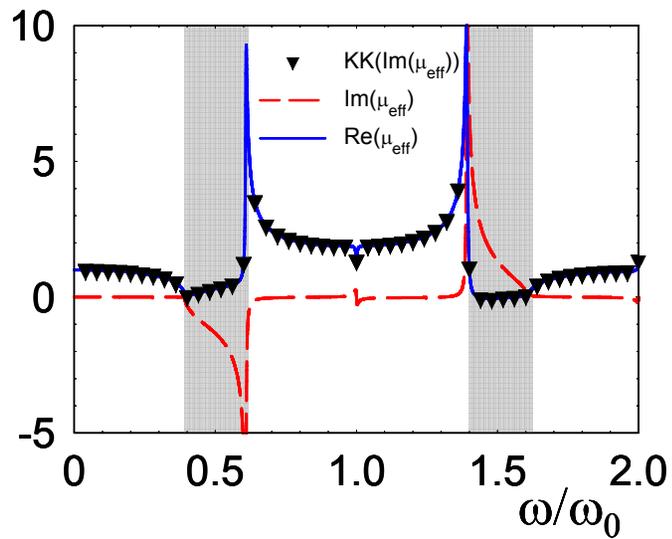

Figure 4:

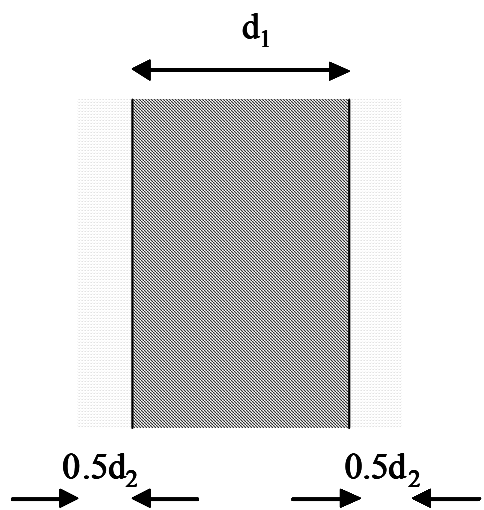
a)

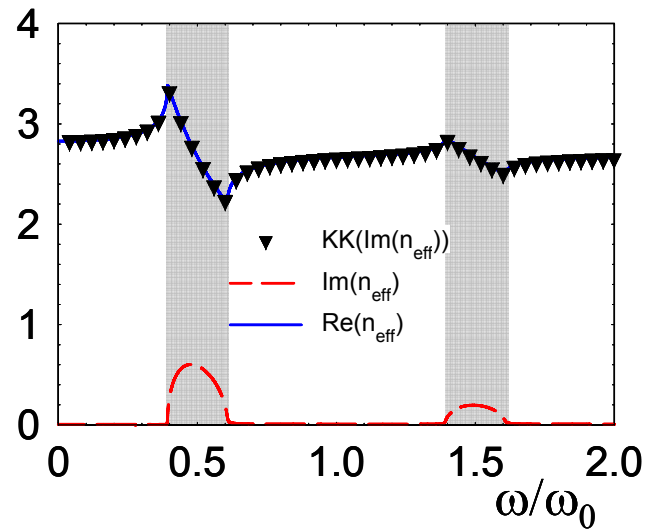
b)

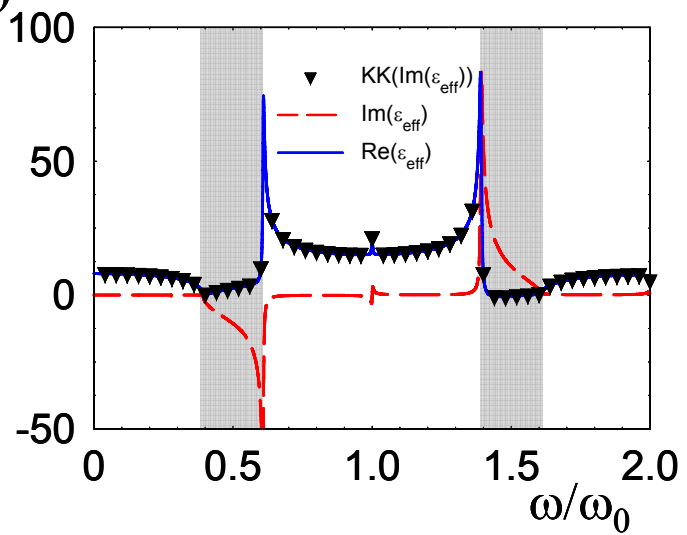
c)

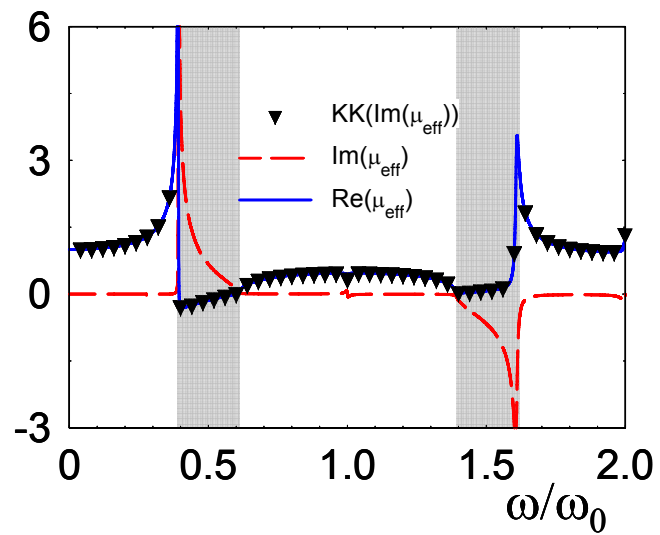
d)

Figure 5:

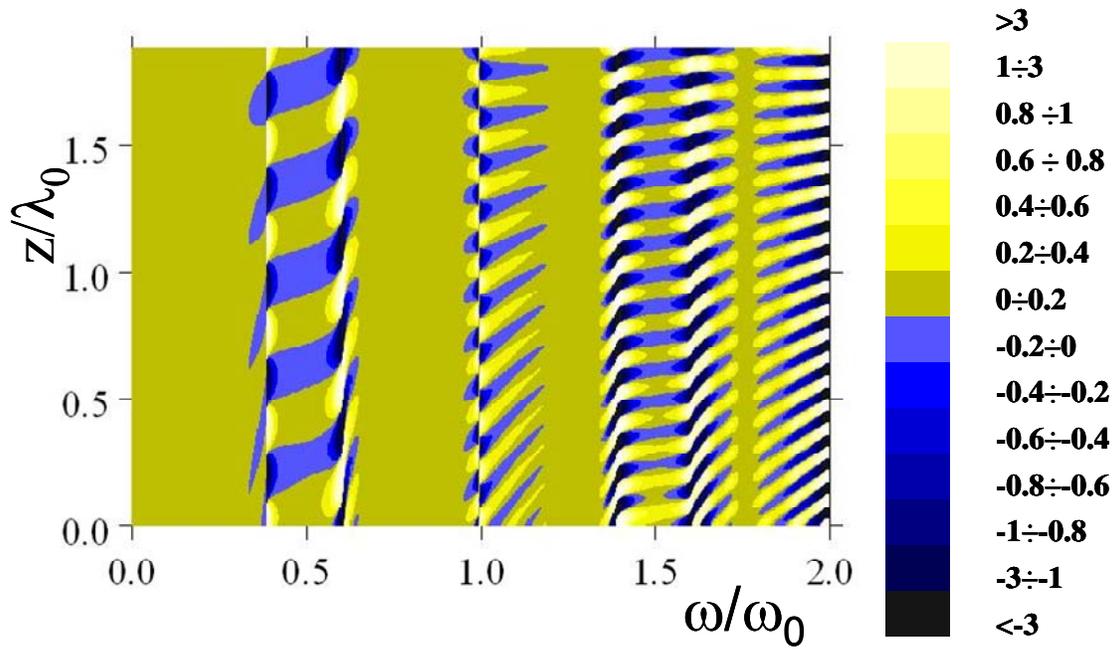

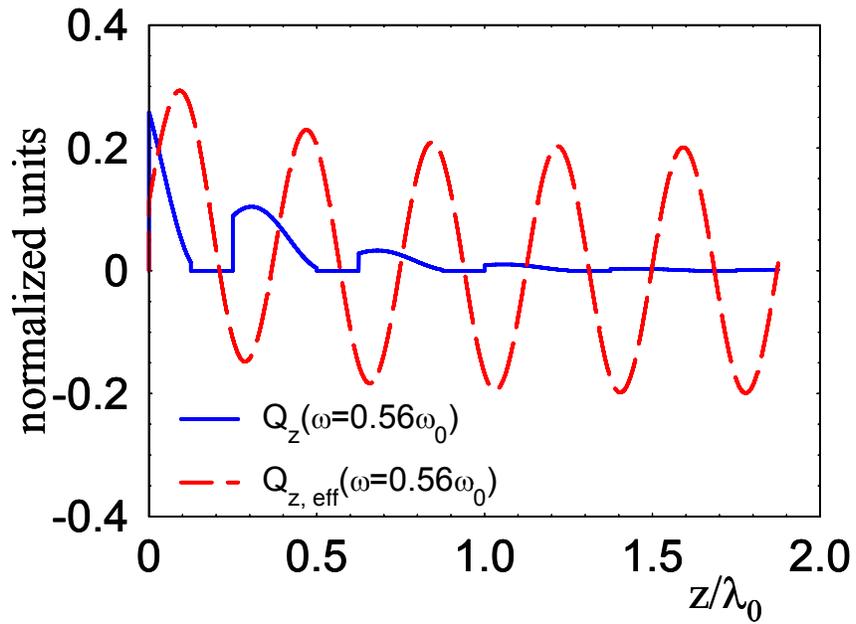

Figure 6:

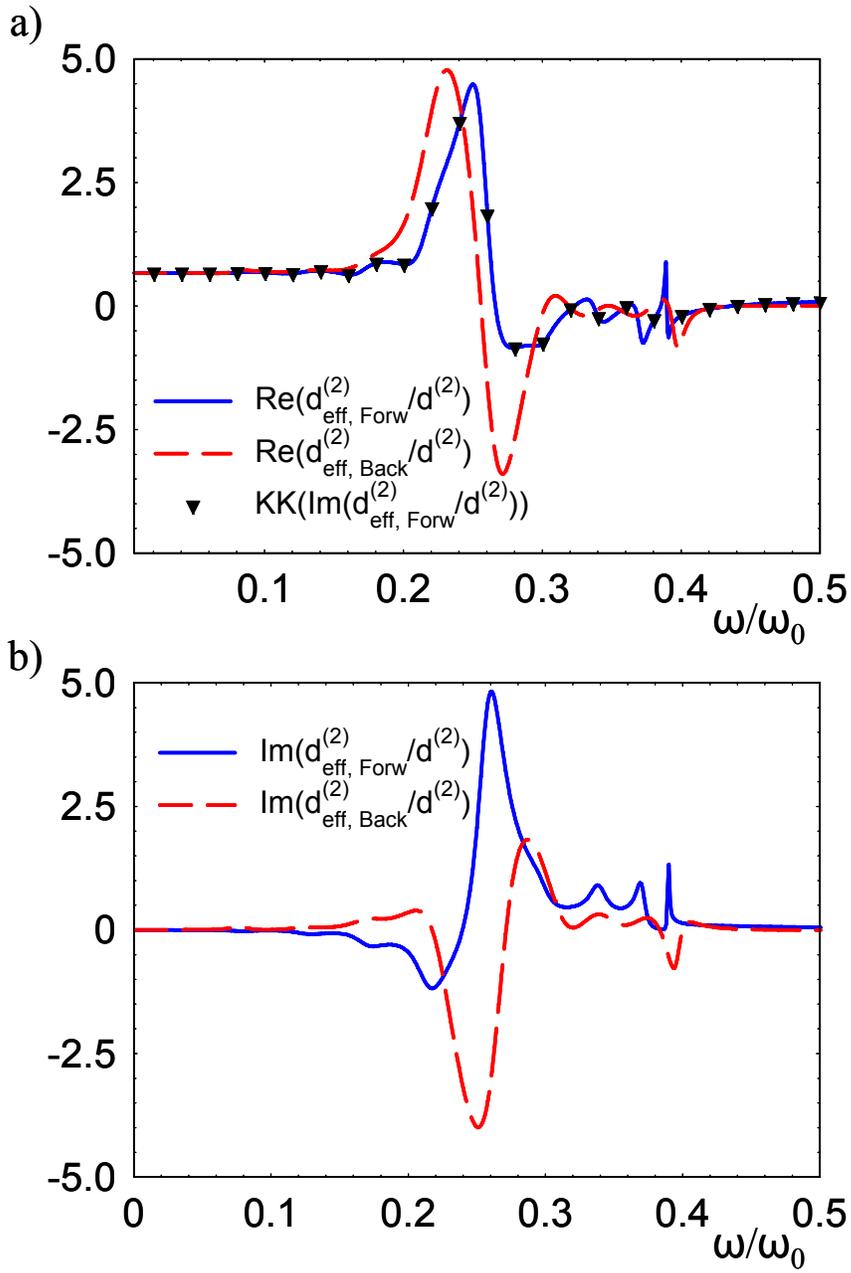